\documentclass[twocolumn,amsmath,amssymb,aps,prl,showpacs,tightenlines]{revtex4}
\usepackage{amsmath}
\usepackage{amssymb}
\usepackage{txfonts}
\usepackage{graphicx}
\usepackage{epstopdf}
\usepackage{dcolumn}
\usepackage{bm}
\usepackage[colorlinks=true,citecolor=blue,linkcolor=magenta]{hyperref}
\begin{document}

\title{Magnon Blockade in a Hybrid Ferromagnet-Superconductor Quantum System}
\author{Zeng-Xing Liu}\email{zengxingliu@hust.edu.cn}
\author{Hao Xiong}\email{haoxiong1217@gmail.com}
\author{Ying Wu}
\affiliation{School of Physics, Huazhong University of Science and Technology, Wuhan 430074, People's Republic of China}
\date{\today}
\begin{abstract}
The implementation of a single magnon level quantum manipulation is one of the fundamental targets in quantum magnetism with a significant practical relevance for precision metrology, quantum information processing, and quantum simulation.
Here, we demonstrate theoretically the feasibility of using a hybrid ferromagnet-superconductor quantum system to prepare a single magnon source based on magnon blockade effects.
By numerically solving the quantum master equation, we show that the second-order correlation function of the magnon mode depends crucially on the relation between the qubit-magnon coupling strength and the driving detuning, and simultaneously signatures of the magnon blockade appear only under quite stringent conditions of a cryogenic temperature.
In addition to providing perception into the quantum phenomena of magnon, the study of magnon blockade effects will help to develop novel technologies for exploring the undiscovered magnon traits at the quantum level and may find applications in designing single magnon emitters.
\end{abstract}

\maketitle

A fundamental type of light-matter interface, magnetic dipole interaction, based on the cavity magnon-microwave photon system has attracted a lot of attention and progressed enormously over the past decade \cite{microwave1,microwave2,microwave3,microwave4,microwave5,microwave6,Bistability1,Bistability2,Access}.
Previous experiments have demonstrated that strong or even ultrastrong coupling between the collective-excitation mode of spin ensembles and microwave photons in the cavity field can be mediated by magnetic dipole interaction because of the extremely high spin density of ferromagnetic materials \cite{microwave1,microwave2,microwave3,microwave4}.
Magnon, regarded as the quantized spin wave, is the dynamic intrinsic excitation of the magnetic ordered body \cite{magnon}, which has been the subject of extensive investigations in many fields of research, for instance, cavity optomagnonics \cite{Optomagnonic1,Optomagnonic2,Optomagnonic3}, hybrid ferromagnetic-superconducting system \cite{qubit1,qubit2,qubit3,Hybrid}, and Dirac or Weyl magnons in topological insulators\cite{Weyl1,Dirac1,Dirac2}.
Many novel phenomena and important applications, ranging from optical cooling of magnon \cite{cooling} and magnon-induced high-order sideband generation \cite{chaos1,chaos2,chaos3} to magnon gradient memory \cite{memory1,memory2} and observation of topological magnon insulator states \cite{Insulator} have been theoretically or experimentally verified.

Recently, the investigation of the quantum characteristics of the magnon-polariton system has fascinated widespread concern and made substantial progress \cite{Bell,Entanglementin1,Entanglementin2,Entanglementin3,Squeezed}.
For example, the generations of magnon-photon-phonon entanglement \cite{Entanglementin1} and squeezed magnon-phonon states \cite{Squeezed} from the cavity magnomechanics, as well as the steady Bell state generation via magnon-photon coupling \cite{Bell} have been proposed.
However, as a typical pure quantum phenomenon, the magnon blockade is still unexplored.
The earliest blockade effect, Coulomb blockade \cite{Coulomb1,Coulomb2,Coulomb3}, was proposed by Gorter et al. \cite{Coulomb} to explain the abnormal increase in resistance of granular metals with temperature.
Subsequently, photon, phonon, and spin blockade effects were gradually discovered in some nonlinear systems \cite{photon1,photon2,phonon1,phonon2,spin1}.
Magnon blockade as a pure quantum effect is one of the most important aspects of exploring quantum properties in the magnon-polariton system.
Furthermore, the investigation of magnon blockade effects is the most critical step in the preparation of single magnon sources, which provides theoretical support for the realization of single magnon level quantum manipulation.


The major emphasis of the present work is to investigate the statistical properties of steady-state magnons based on a hybrid ferromagnet-superconductor quantum system \cite{qubit1,qubit2,qubit3,Hybrid} and discuss the magnon blockade effects under the current experimental conditions by evaluating the equal-time second-order correlation function $g^{(2)}(0)$ of the magnon mode.
Furthermore, by numerically solving the quantum master equation for the density matrix \cite{mast1}, we observed the classical bunching effect $g^{(2)}(0)>1$ and nonclassical anti-bunching effect $g^{(2)}(0)<1$ of magnons \cite{photon2,phonon1,phonon2}, and particularly, for $g^{(2)}(0)\rightarrow 0$ indicates magnon blockade where strong interactions between the magnon and the qubit prevent the excitation of multiple magnons at the same time.
In addition, we have shown that magnon blockade effects exhibit a high dependence on the qubit-magnon interaction strength as well as the driving detuning, and the influences of the thermal noise of the system environment on the magnon blockade effects have been discussed in detail.
Besides their fundamental scientific significance, the exploration of magnon blockade is crucial for the study of magnon at the quantum level and opens up a pathway for designing single magnon emitter \cite{single1,single2}.
Our results, therefore, provide theoretical possibilities for quantum manipulation at the single-magnon level.


\begin{figure}[htbp]
\centering
\includegraphics [width=1\linewidth] {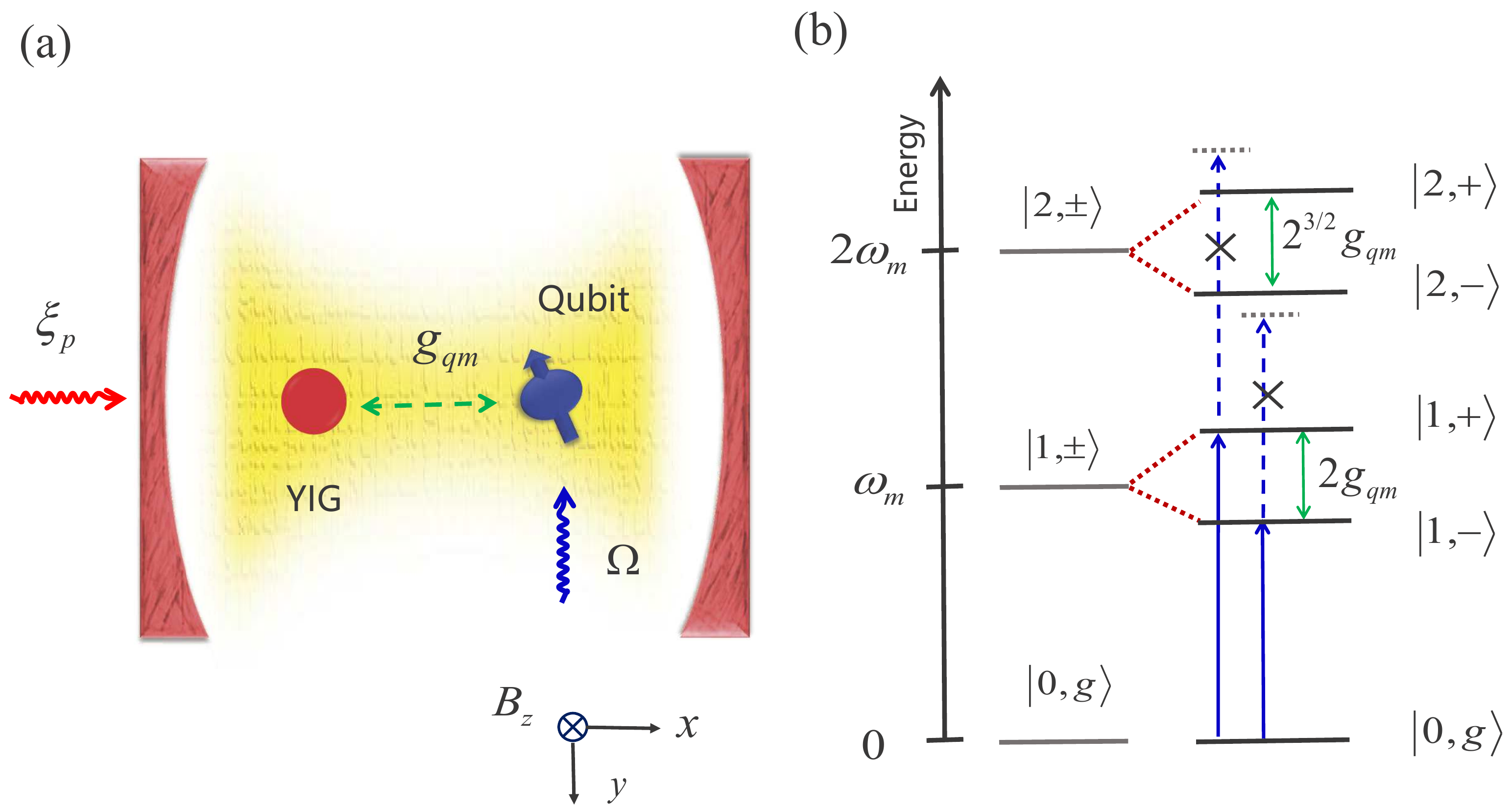}
\caption{(a) Schematic illustration of the hybrid ferromagnet-superconductor quantum model, in which a single-crystalline ferromagnetic YIG sphere and a transmontype superconducting qubit are installed in a microwave cavity \cite{qubit2}.
The qubit is driven by a monochromatic drive field (with strength $\Omega$) and a weak probe field (with strength $\xi_{p}$) is applied into the YIG sphere to observed the magnon blockade effects.
The qubit and magnons can be effectively coupled via the exchange of virtual cavity photons (with the coupling strength $g_{qm}$).
A static magnetic field $B_{z}$ aligned along the hard magnetization axis $[100]$ of the YIG sphere can be used to adjust the frequency of the Kittel (magnon) mode.
(b) Energy levels of the qubit-magnon quantum system with two-magnon state involved for the Kittel mode. The coupling between the magnon mode and the qubit induces the vacuum Rabi splitting of the dressed states.}
\label{fig:1}
\end{figure}

The physical model in our studies is a hybrid ferromagnet-superconductor quantum system (shown in Fig. \ref{fig:1}(a)) that combines two collective-excitation modes: i.e., the magnon mode in a ferromagnetic yttrium iron garnet ($\rm{Y_{3}Fe_{5}O_{12}}$ YIG) sphere and a transmontype superconducting qubit \cite{qubit2}.
The microwave cavity made of aluminum and oxygen-free copper (the YIG sphere is placed in the copper part of the cavity while the qubit is fixed in the aluminum part of the cavity) is positioned in a dilution refrigerator at a base cryogenic temperature of $\sim$ 20 mK.
In this circumstance, the aluminum is superconducting and diamagnetic, namely, the qubit is immune to the external magnetic field, whereas the YIG sphere is non-superconducting and its frequency is magnetically dependent \cite{qubit2}.
The magnon mode and the qubit can, respectively, be strongly coupled to the microwave cavity field when their frequencies resonate with the microwave field.
Here, we need to point out that the coupling between the magnon mode and superconducting qubit can also be achieved in a simpler cavity design, as Ref. \cite{qubit1} has implemented.
Experimentally, the coupling strength between the magnon (qubit) and the microwave field can be well characterized by controlling the far-detuning between the qubit (magnon) and the microwave field frequencies \cite{qubit1,qubit2}.
In addition, when the frequency of the magnon mode is tuned to be nearly resonant with the transition frequency of the qubit, the magnon mode and the qubit can achieve effective coupling and the coherent exchange between the qubit excitation and the magnon is mediated by the virtual-photon excitation in the microwave cavity mode \cite{qubit1,qubit2,qubit3}.
In this context, the interaction between the magnon mode and the superconducting qubit can be well described by a Jaynes-Cummings-type \cite{JC} Hamiltonian (in the rotating wave approximation) ${\rm{H_{int}}} = \hbar g_{qm}(\sigma_{+}m + \sigma_{-}m^{\dagger})$,
where $g_{qm} = g_{q}g_{m}/\Delta$ is the effective qubit-magnon coupling strength with $g_{q}$ $(g_{m})$ the qubit (magnon)-photon coupling strength \cite{qubit1,qubit2} and $\Delta$ the frequency difference between the qubit (magnon) and the microwave cavity mode. $\sigma_{-}$ ($\sigma_{+} = (\sigma_{-})^{\dagger}$) and $m$ ($m^{\dagger}$) are the annihilation (creation) operators of the qubit excitation and the magnon, respectively.
The coupling between the magnon mode and the qubit gives rise to the vacuum Rabi splitting of the dressed states (shown in Fig. \ref{fig:1}(b)).

In order to implement and dynamically manipulate the magnon blockade, we use a monochromatic control field with the frequency $\omega_{d}$ to drive the superconducting qubit and the Hamiltonian is ${\rm{H}}_{\rm{d}} = \hbar \Omega(\sigma_{+}e^{-i\omega_{d}t} + \sigma_{-}e^{i\omega_{d}t})$, where $\Omega = \mathbb{K}\sqrt{\rm{P}_{d}}$ with the drive parameter $\mathbb{K}$ = 103 $\rm{MHz/mW^{1/2}}$ \cite{qubit2} and the drive power $\rm{P}_{d}$, is the coupling strength between the control field and the qubit.
Furthermore, for the observation of magnon blockade effects, we consider a weak probe field with the Hamiltonian ${\rm{H}}_{\rm{p}} = \hbar \xi_{p}(m^{\dagger}e^{-i\omega_{d}t} + me^{i\omega_{d}t})$ (with frequency $\omega_{p}$ and strength $\xi_{p}$) applied into the YIG sphere.
The total Hamiltonian of the driven qubit-magnon hybrid quantum system, therefore, is given by (we set $\hbar$ = 1)
\begin{eqnarray}\label{equ:01}
  \rm{H_{tot}} &=& \frac{1}{2}\omega_{q}\sigma_{z}+\omega_{m}m^{\dagger}m+g_{qm}(\sigma_{+}m + \sigma_{-}m^{\dagger}) \\
  &&+\Omega(\sigma_{+}e^{-i\omega_{d}t}+\sigma_{-}e^{i\omega_{d}t})+\xi_{p}(m^{\dagger}e^{-i\omega_{p}t} + me^{i\omega_{p}t})\nonumber,
\end{eqnarray}
where $\sigma_{z}$ is the Pauli operator for the qubit, and $\omega_{q}$ ($\omega_{m}$) denotes the frequency of the qubit (magnon) mode.
The time-dependence can be removed to simplify the simulation by a rotating frame transformation with respect to the control field frequency $\omega_{d}$, and consequently, the effective Hamiltonian of this hybrid qubit-magnon system can be obtained as
\begin{eqnarray}\label{equ:02}
  \rm{H_{eff}} &=& \frac{1}{2}\Delta_{q}\sigma_{z}+\Delta_{m}m^{\dagger}m+g_{qm}(\sigma_{+}m + \sigma_{-}m^{\dagger})\nonumber \\
  &&+\Omega(\sigma_{+}+\sigma_{-})+\xi_{p}(m^{\dagger}e^{-i\delta t}+me^{i\delta t}),
\end{eqnarray}
where $\Delta_{q} \equiv \omega_{q} - \omega_{d}$ ($\Delta_{m} \equiv \omega_{m} - \omega_{d}$) is the frequency detuning between the qubit (magnon) and the control field.
Here, for simplicity, we assume that the beat frequency between the control field and the probe field is $\delta \equiv\omega_{d} - \omega_{p} = 0$, and simultaneously $\Delta = \Delta_{q} = \Delta_{m}$ for $\omega_{q} \simeq \omega_{m}$.

After taking into the damping caused by the system-bath coupling, the dissipative dynamical evolution of the hybrid qubit-magnon system in terms of the magnon mode is governed by the master equation \cite{mast1}
\begin{eqnarray}\label{equ:03}
  \frac{\partial}{\partial t}\rho &=& -i[\rm{H_{eff}}, \rho]+\frac{\kappa_{m}}{2}({n}_{th}+1)\mathcal{L}_{m}[\rho]\nonumber \\
  &&+\frac{\kappa_{m}}{2}\rm{{n}_{th}}\mathcal{L}_{m^{\dagger}}[\rho]
  +\frac{\kappa_{q}}{2}\mathcal{L}_{\sigma_{-}}[\rho],
\end{eqnarray}
where $\rho$ is the system density matrix and
\begin{eqnarray}\label{equ:04}
\mathcal{L}_{m}[\rho] &=& (2m\rho m^{\dagger} - m^{\dagger}m\rho - \rho m^{\dagger}m),\nonumber \\
  \mathcal{L}_{\sigma_{-}}[\rho] &=& (2\sigma_{-}\rho \sigma_{+} - \sigma_{+}\sigma_{-}\rho - \rho \sigma_{+}\sigma_{-}),
\end{eqnarray}
denote the Lindbland superoperators for the operator $m$ and $\sigma_{-}$, respectively \cite{mast2}.
$\kappa_{m}$ and $\kappa_{q}$ are, respectively, the dissipation rates of the magnon mode and the qubit, and $\rm{{n}_{th}} =[{\rm{exp}}(\frac{\hbar\omega_{m}}{\rm{K_{\rm{B}}T}})-1]^{-1}$ is the equilibrium thermal magnon occupation number, with the Boltzmann constant $\rm{K_{B}}$ and the ambient temperature $\rm{T}$.
Analogous to the quantum statistical properties of photon (phonon) \cite{photon1,photon2,phonon1}, the properties of steady-state magnon statistics can also be well characterized by using the equal-time second-order correlation function \cite{mast1}
\begin{eqnarray}\label{equ:05}
  g^{(2)}(0) = \frac{{\rm{Tr}}(\rho m^{\dagger}m^{\dagger}mm)}{[{\rm{Tr}}(\rho m^{\dagger}m)]^{2}} =\frac{\langle m^{\dagger}m^{\dagger}mm\rangle}{\langle m^{\dagger}m\rangle^{2}},
\end{eqnarray}
which can be obtained by numerically solving the master equation (the Monte Carlo wave-function simulation \cite{mast3}) in Eq. (\ref{equ:03}) with the effective Hamiltonian $\rm{H_{eff}}$ in Eq. (\ref{equ:02}).
The value of the second-order correlation function $g^{(2)}(0) > 1$ implies that the steady-state magnon number distribution satisfies the super-Poissonian statistics, which is referred to as the magnon bunching effect.
If, on the other hand, $0 < g^{(2)}(0) < 1$, the distribution of magnon number satisfies the sub-Poissonian statistics referred to as the magnon anti-bunching effect \cite{photon2,phonon1,phonon2}.
In particular, $g^{(2)}(0)\rightarrow 0$ indicates that the occurrence of the magnon blockade.
In our scheme, the second-order correlation function can reach as low as $g^{(2)}(0) \rightarrow 10^{-5}$ under the current experimental conditions \cite{qubit1,qubit2}, which is a good signature of magnon blockade.
Our studies, therefore, provide theoretical support for the preparation of a single photon source that can be applied to both fundamental problems in quantum magnonic as well as tasks of magnon quantum simulation and hybrid quantum devices \cite{nanomagnet1,nanomagnet2}.

\begin{figure}[htbp]
\centering
\includegraphics [width=1\linewidth] {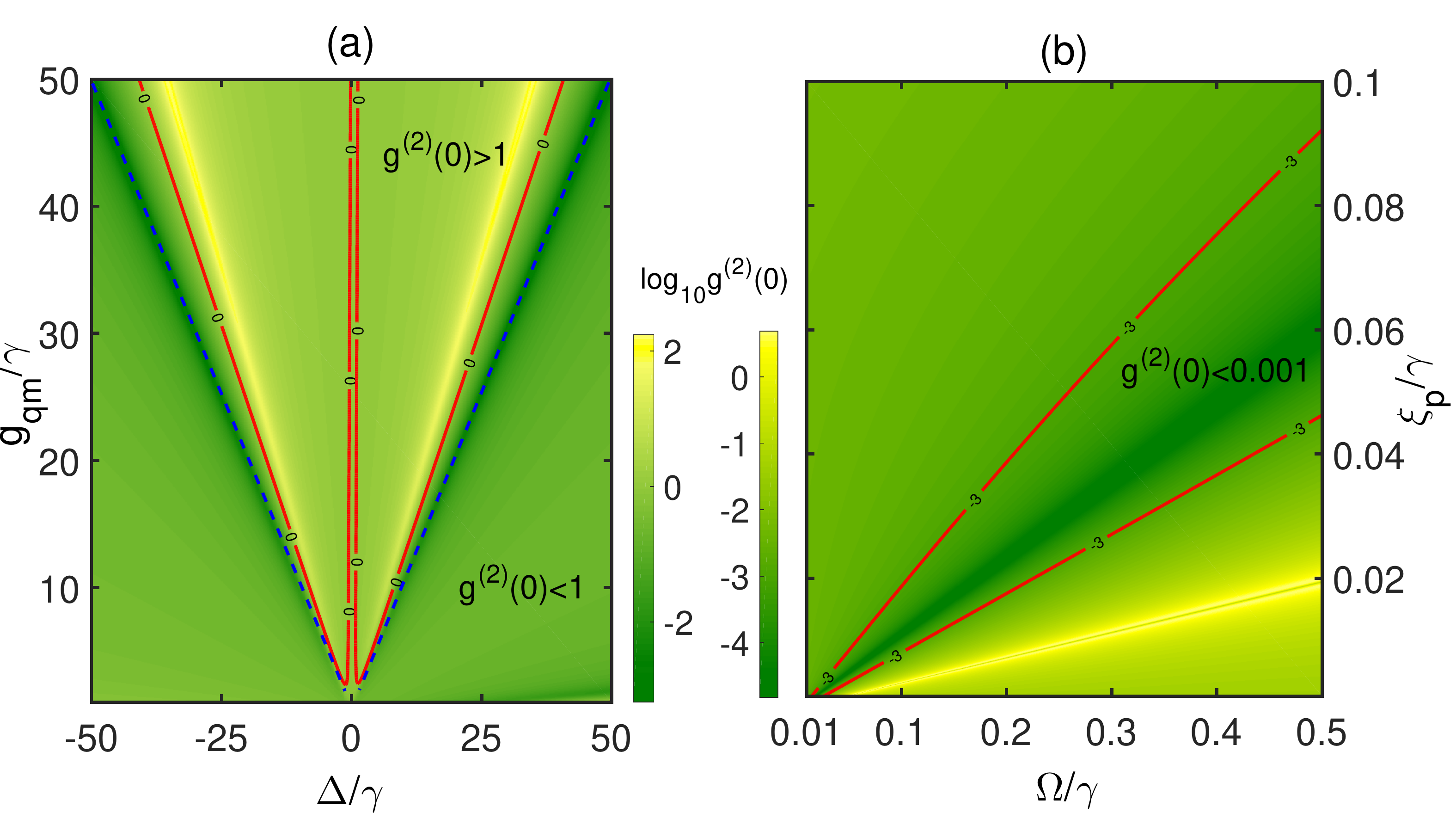}
\caption{The steady-state equal-time second-order correlation function $\rm{log}_{10}g^{(2)}(0)$ is plotted varies with (a) the driving detuning $\Delta/\gamma$ (here, for convenience, $\gamma=2\pi\times1\rm{MHz}$ used to scale the frequencies throughout the article) and the effective qubit-magnon coupling strength $g_{qm}/\gamma$, (b) the control field strength $\Omega/\gamma$ and the probe field strength $\xi_{p}/\gamma$, respectively.
We use $\Omega/\gamma = 0.1$ ($\rm{P}_{d} \simeq$ 0.037 $\mu \rm{W}$) and $\xi_{p}/\gamma = 0.001$ (the probe power $\rm{P}_{p} \simeq$ 0.0037 $\rm{nW}$) in (a), and $\Delta/\gamma = 21$ and $g/\gamma = 21$ in (b).
The other system parameters we take are $\kappa_{m}/\gamma=1.4$, $\kappa_{q}/\gamma=1.2$, and $\rm{{n}_{th}} = 0$.}
\label{fig:2}
\end{figure}

In the following, we discuss the dependence of the equal-time second-order correlation function on the system parameters, for instance, the driving detuning, the effective qubit-magnon coupling strength, and the thermal magnon occupation number, et cetera.
In Fig. \ref{fig:2}(a), concretely, we plot the second-order correlation function $g^{(2)}(0)$ as a function of the driving detuning $\Delta/\gamma$ and the effective qubit-magnon coupling strength $g_{qm}/\gamma$.
The red contour indicates that the value of the second-order correlation function $\rm{log}_{10}g^{(2)}(0) = 0$,  which is a clear boundary that separates the classical regime $g^{(2)}(0)>1$ from the regime of pure quantum correlations $g^{(2)}(0)<1$ \cite{photon2}.
Namely, within the red contour, the magnon exhibits a bunching effect, while outside the red contour, the magnon exhibits an anti-bunching effect which is a pure quantum effect.
In particular, when the value of the driving detuning is approximately equal to the coupling strength $\Delta/\gamma \simeq \pm g_{qm}/\gamma$,
the second-order correlation function $g^{(2)}(0)$ is the smallest, as the blue dotted line shown, which is the optimal parameter condition for investigating the quantum properties of the magnons.
Furthermore, controlling the magnon blockade effect by adjusting the external incident field is the most experimentally efficient method.
The relationships between the second-order correlation function $g^{(2)}(0)$ and the control field strength $\Omega/\gamma$ as well as the probe field strength $\xi_{p}/\gamma$ are plotted in Fig. \ref{fig:2}(b).
An optimal region where the value of the second-order correlation function $g^{(2)}(0) < 0.001$ has been obtained, which reminds us of the feasibility of realizing magnon blockade by properly adjusting the external incident fields.
The physical mechanism of the magnon blockade can be intuitively understood from the energy-level diagram in Fig. \ref{fig:1}(b), where $|n, \pm\rangle$ ($n = 0, 1 ,2$) are the dressed states for the coupled qubit-magnon system. $|n\rangle$ and $|g\rangle$ are, respectively, the eigenstates of the magnon number and the ground state of the qubit (refer as an effective two-level system).
Here, magnon blockade indicates that magnons can only be excited individually, but not two or more magnons can be excited together. In the energy level picture in Fig. \ref{fig:1}(b), correspondingly, the transitions between the quantum states $|0, g\rangle$ and $|1, \pm\rangle$ are allowed, namely, the condition of the drive field detuning $\omega_{d} - \omega_{m} = \pm \Delta$ ($\Delta = g_{qm}$) must be satisfied.
And simultaneously, the transition $|1, \pm\rangle$ $\rightarrow$ $|2, \pm\rangle$ are suppressed due to the qubit-induced vacuum Rabi splitting, viz, $2\omega_{m} \pm 2\Delta \neq 2\omega_{m} \pm \frac{2^{3/2}}{2}g_{qm}$.
In addition, to observe the magnon blockade, the interaction between the qubit and the magnon mode should work in the strong-coupling regime, namely, $g_{qm} \gg \kappa_{q}, \kappa_{m}$, which is consistent in our numerical simulation and has been experimentally demonstrated \cite{qubit1,qubit2,qubit3}.

\begin{figure}[htbp]
\centering
\includegraphics [width=1\linewidth] {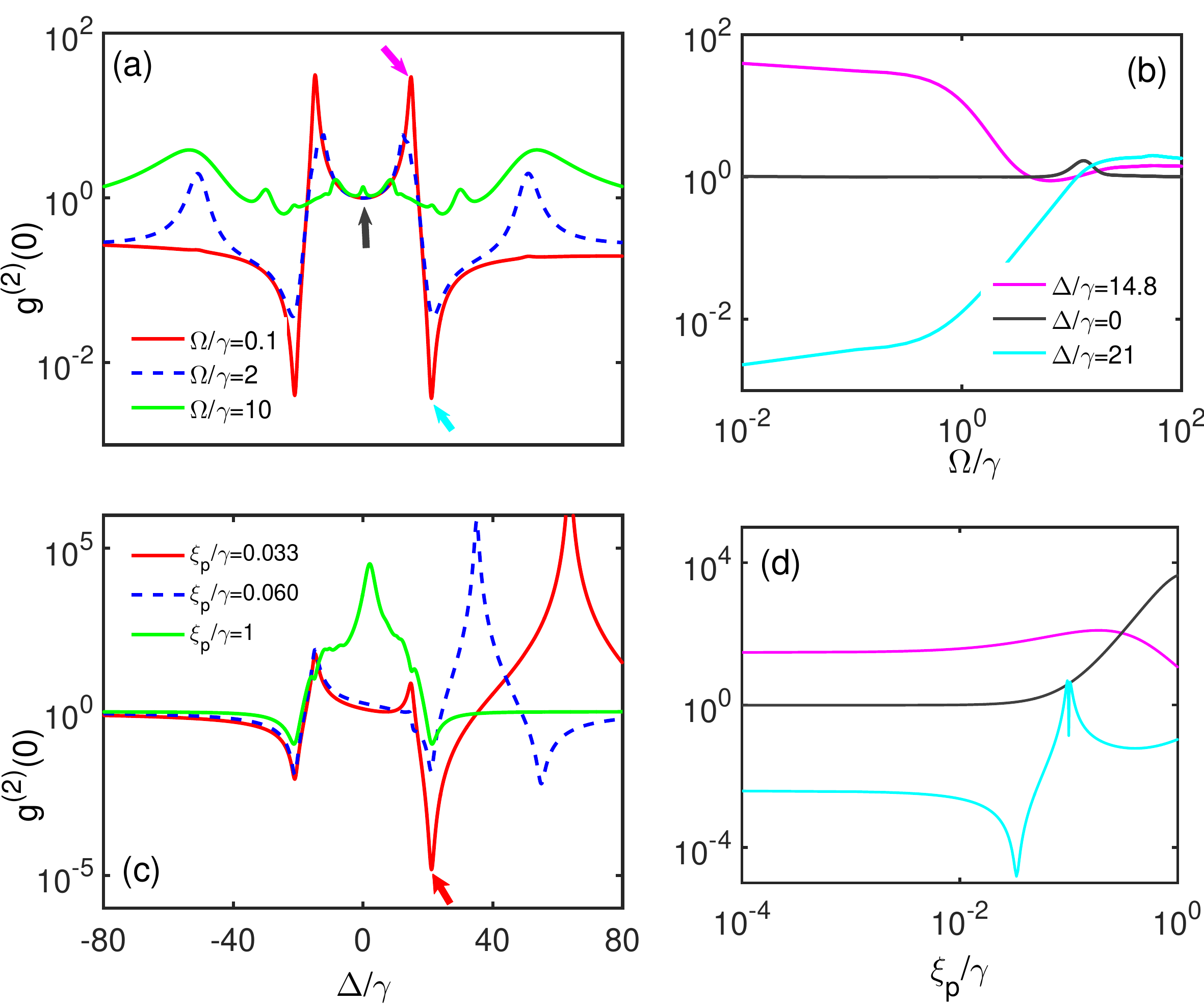}
\caption{The equal-time second-order correlation function $g^{(2)}(0)$ versus the detuning $\Delta/\gamma$ under the different values of the control field strength $\Omega/\gamma = (0.1, 2, 10)$ in (a) and the probe field strength $\xi_{p}/\gamma = (0.033, 0.06, 1)$ in (c).
The dependence of $g^{(2)}(0)$ on $\Omega/\gamma$ and $\xi_{p}/\gamma$ are plotted in (b) and (d) for $\Delta/\gamma = (14.8, 0, 21)$, respectively.
We use $\xi_{p}/\gamma = 0.001$ in (a) and (b), $\Omega/\gamma = 0.1$ in (c) and (d), and the other parameters are the same as those in Fig. \ref{fig:2}(b).}
\label{fig:4}
\end{figure}

Next, we observe a high dependence of the second-order correlation function $g^{(2)}(0)$ on the driving detuning and more clear results are shown in Fig. \ref{fig:4}.
Specifically, we plot the second-order correlation function $g^{(2)}(0)$ as a function of the driving detuning
$\Delta/\gamma$ when the control and probe fields strength, respectively, take the values of $\Omega/\gamma = (0.1, 2, 10)$ and $\xi_{p}/\gamma = (0.033, 0.06, 1)$ in Figs. \ref{fig:4}(a) and (c).
As expected, the correlation function $g^{(2)}(0)$ manifests a strong anti-bunching effect at $\Delta/\gamma \simeq \pm 21$, and simultaneously, reveals a strong bunching effect at $\Delta/\gamma \simeq \pm 14.8$.
Physically, as shown by the pink arrow in Fig. \ref{fig:4}(a) ($\Delta/\gamma \simeq 14.8$), the transition between $|1, \pm\rangle$ and $|2, \pm\rangle$ is enhanced because the transition condition $2\Delta \simeq \frac{2^{3/2}}{2}g_{qm} \simeq 29.7$ for $g_{qm}/\gamma = 21$ is satisfied.
In this context, the magnon tends to appear in pairs (bunching effect).
When the driving detuning $\Delta/\gamma \simeq 21$, as shown by the sky-blue arrow in Fig. \ref{fig:4}(a), the transition $|1, \pm\rangle$ $\rightarrow$ $|2, \pm\rangle$, however, is inhibited owing to $2\Delta \neq \frac{2^{3/2}}{2}g_{qm}$, which is the physical origin of the magnon blockade.
In particular, when we choose the probe field strength $\xi_{p}/\gamma = 0.033$ (shown by the red line in Fig. \ref{fig:4}(c)), the second-order correlation function occurs an extremely low value $g^{(2)}(0)$ $\rightarrow$ $10^{-5}$ (the red arrow), which is an excellent signal of magnon blockade.
The quantitative relationship for magnon blockade will be more intuitive by displaying the correlation function $g^{(2)}(0)$ as a function of the control field strength $\Omega/\gamma$ and the probe field strength $\xi_{p}/\gamma$ for a given value of $\Delta/\gamma = (14.8, 0, 21)$ in Figs. \ref{fig:4}(b) and (d), respectively.
We can obviously see that with the increase of the control field strength, both the anti-bunching effect and the bunching effect of the magnon are destroyed until they completely disappear ($g^{(2)}(0) \rightarrow 1$).
The above results demonstrate that our proposal provides a prospective approach for performing the manipulation of magnon using qubit in the quantum domain and the implementation of magnon blockade with current experimental technologies \cite{qubit1,qubit2,qubit3}.

Finally, we discuss how does the ambient thermal noise affects the magnon blockade?
\begin{figure}[htbp]
\centering
\includegraphics [width=0.95\linewidth] {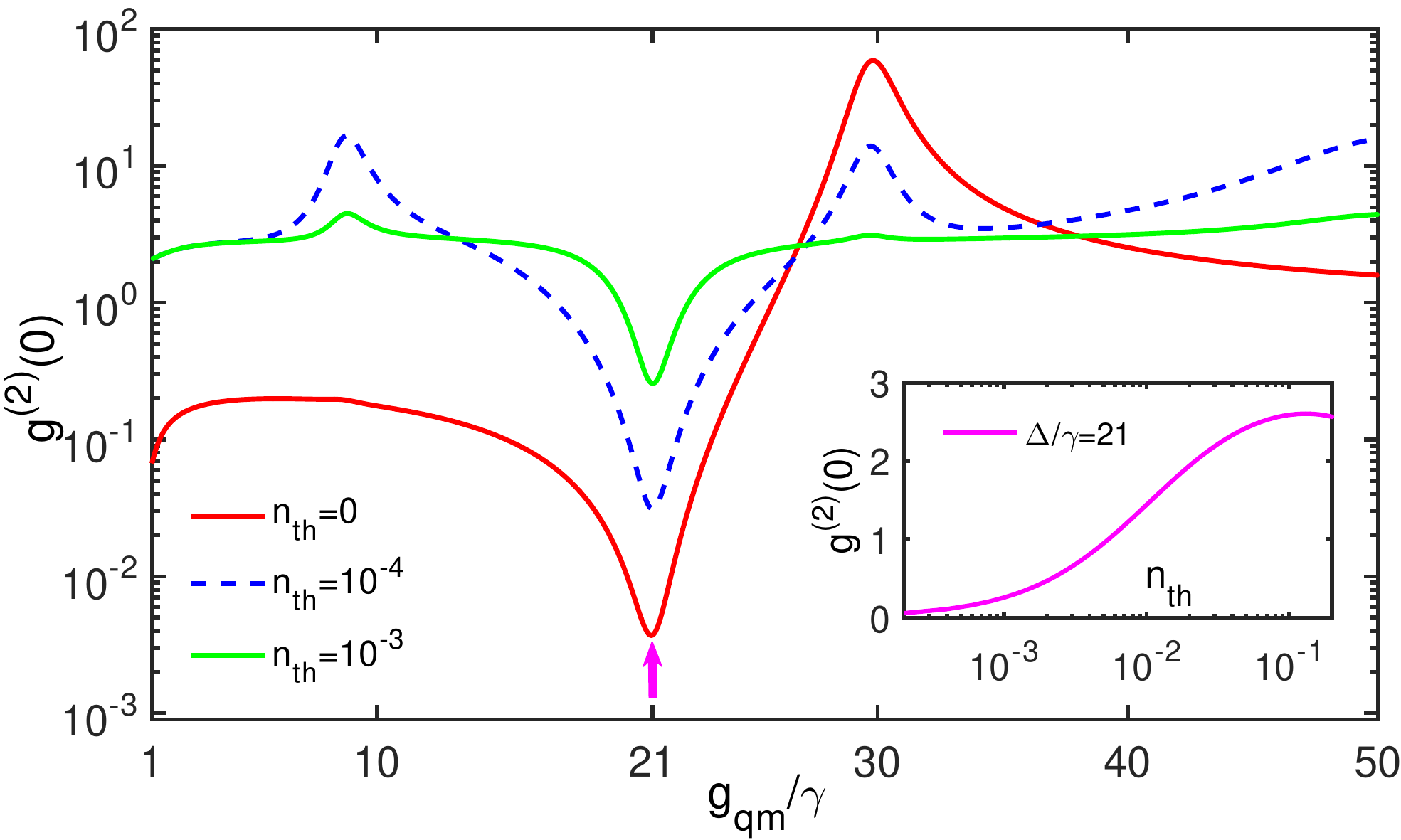}
\caption{The equal-time second-order correlation function $g^{(2)}(0)$ as a function of the effective qubit-magnon coupling strength $g_{qm}/\gamma$ under the different thermal magnon occupation number $\rm{{n}_{th}} = (0, 10^{-4}, 10^{-3})$.
We use $(\Delta, \Omega, \xi_{p})/\gamma$ = (21, 0.1, 0.001), and the other parameters are the same as those in Fig. \ref{fig:2}.}
\label{fig:5}
\end{figure}
To answer this question, in Fig. \ref{fig:5}, we plot the equal-time second-order correlation function $g^{(2)}(0)$ varies with the qubit-magnon coupling strength $g_{qm}/\gamma$ under the different thermal magnon occupation number $\rm{{n}_{th}} = (0, 10^{-4}, 10^{-3})$ by numerically solving the master equation in Eq. (\ref{equ:03}) in the case of $\Delta/\gamma = 21$.
It is clear that the thermal magnon number has a significant impact on the magnon blockade effect.
In the present work, the hybrid qubit-magnon system is placed in a dilution refrigerator at a base cryogenic temperature of $\sim$ 20 mK \cite{qubit2} and the thermal magnon occupation number $\rm{{n}_{th}} =[{\rm{exp}}(\frac{\hbar\omega_{m}}{\rm{K_{\rm{B}}T}})-1]^{-1}$ is evaluated to be $\rm{{n}_{th}} \simeq 10^{-9}$ for the magnon frequency $\omega_{m}/2\pi = 8.5 \rm{GHz}$, therefore, the hereinabove discussions about the magnon blockade is experimentally feasible.
When we fine-tune the ambient temperature to T $\sim$ 45 mK and T $\sim$ 60 mK, as the blue dotted and green solid lines are shown, the magnon blockade effects undermine rapidly with the increase of the thermal magnon occupation number $\rm{n_{th}} \sim 10^{-4}$ to $\rm{n_{th}} \sim 10^{-3}$.
More clear results are shown from the illustration in Fig. \ref{fig:5}, in which the correlation function $g^{(2)}(0)$ varies with the thermal magnon occupation number under the circumstance $\Delta/\gamma = 21$ (the pink arrow) are plotted.
We can see that when the ambient temperature is T $\sim$ 100 mK, i.e.,  the thermal magnon number $\rm{{n}_{th}} \sim 10^{-2}$, the magnon blocking effect, or more precisely, the magnon anti-bunching effect, disappears completely.
This signifies that the magnon blockade is limited by the thermal noise of the system environment, and thus, to better observe the desired magnon blockade, the YIG sphere must be cooled to a low-temperature environment.

We propose how to implement and manipulate the magnon blockade effect in a hybrid qubit-magnon quantum system, in which the strong interaction between the magnon and the qubit induces the vacuum Rabi splitting.
By numerically solving the quantum master equation with the system Hamiltonian, we analyze the magnon statistical properties characterized by the second-order correlation function of the magnon mode and observe the magnon blockade effect.
The dependence of the magnon blockade effect on the relevant system parameters $\Delta$, $g_{qm}$, and $\Omega$ have been studied in detail, and we find that the magnon blockade effect is limited by the thermal magnon occupation number $\rm{n_{th}}$.
Within the reach of experimental capabilities, we believe that the observation of magnon blockade would provide the fundamental ingredient for potential applications of the hybrid qubit-magnon quantum system as a single magnon emitter, which may have potentially significant applications in quantum simulation, quantum information processing, and even hybrid quantum networks.

The work was supported by the National Key Research and Development Program of China (Grant No. 2016YFA0301203), the National Science Foundation of China (Grant Nos. 11774113, 11405061), and the Fundamental Research Funds for the Central Universities (Grant No. 2019kfyRCPY111).

\end{document}